\documentclass[article]{aa}
\usepackage{txfonts}
\usepackage{natbib}
\usepackage{subfig}
\usepackage{deluxetable}
\bibpunct{(}{)}{;}{a}{}{,} % to follow the A&A style
\usepackage{graphicx}

\begin{document}

\title{The Opacity of Spiral Galaxy Disks V:\\
dust opacity, HI distributions and sub-mm emission.
\thanks{Research support by NASA 
through grant number HST-AR-08360 from the Space Telescope Science 
Institute (STScI), the STScI Discretionary Fund (grant numbers 82206 and 82304) and the Kapteyn 
Institute of Groningen University.}
}
\author{B. W. Holwerda \inst{1,2} \and R. A. Gonz\'alez \inst{3} \and Ronald J. Allen \inst{1} \and P. C. van der Kruit \inst{2}}

\offprints{B.W. Holwerda, \email{Holwerda@stsci.edu}}

\institute{Space Telescope Science Institute, Baltimore, MD 21218
\and
Kapteyn Astronomical Institute, University of Groningen, PO Box 800, 9700 AV Groningen, the Netherlands.
\and
Centro de Radiastronom\'{\i}a y Astrof\'{\i}sica, Universidad Nacional Aut\'{o}noma de M\'{e}xico, 58190 Morelia, Michoac\'{a}n, Mexico}

\date{Received / Accepted}

\titlerunning{dust emission, extinction and HI}
\authorrunning{Holwerda et al.}

\abstract{
The opacity of spiral galaxy disks, from counts of distant galaxies, is compared 
to HI column densities. The opacity measurements are calibrated using the 
``Synthetic Field Method'' from \cite{Gonzalez98,Holwerda05a}.
% Conclusion 1
When compared for individual disks, the HI column density and dust opacity 
do not seem to be correlated as HI and opacity follow different radial profiles.
% Conclusion 2
To improve statistics, an average radial opacity profile is compared to an average 
HI profile. Compared to dust-to-HI estimates from the literature, more extinction is 
found in this profile. This difference may be accounted for by an underestimate of 
the dust in earlier measurements due to their dependence on dust temperature.
Since the SFM is insensitive to the dust temperature, the ratio between the SFM 
opacity and HI could very well be indicative of the true ratio. 
% Conclusion 3
Earlier claims for a radially extended cold dust disk were based on sub-mm observations. 
A comparison between sub-mm observations and counts of distant galaxies is therefore 
desirable. We present the best current example of such a comparison, M51, for which 
the measurements seem to agree. However, this remains an area where improved 
counts of distant galaxies, sub-mm observations and our understanding of dust 
emissivity are needed.
}

\maketitle

\section{\label{secHIintro}Introduction}

The relationship between gas and dust in spiral disks has been the focus of many 
observational studies often combined with efforts to characterize the chemical 
composition. Dust plays an important role in the energy and chemistry budgets 
of a disk, as is evident in our own Galaxy. The question comes simply down to 
whether the dust is distributed as the stars, which produce it, or like the gas, such 
as atomic hydrogen, which is dynamically coupled to it. Furthermore, some fraction 
of the gas which is associated with the dust will be in the form of cold dark molecular 
clouds, which may be of very high opacity. Cold dust can in principle be detected by 
either sub-millimeter emission or through the extinction of a background source.

% dark clouds and HI column densities
A first attempt to characterize cold dust clouds and their relation to HI in a nearby 
galaxy, was by \cite{Hodge80}; his figures 8 and 9 show the radial distribution of 
the number of dark clouds and HI column density along the minor and major axes 
of M31. No correlation between the two tracers was found.

In recent years, the Infrared Space Observatory (ISO), and the Sub-Millimeter 
Common User Bolometer Array (SCUBA) on the James Clerk Maxwell Telescope 
have detected and mapped the cold dust component of spiral disks. This cold 
component has been found at much larger radii in galaxy disks than the warm dust.
\cite{Alton98a} and \cite{Trewhella00} found evidence from ISO for cold dust at 
larger radii. \cite{Alton98b,Alton00b} studied the distribution of emission in 
NGC 891 and also found evidence for a cold dust disk.
\cite{Alton00c} found evidence for dust outside the optical disk in NGC 660. 
In addition, \cite{Bianchi00b} and \cite{Alton04} found a correlation between 
CO and 850 micron emission in NGC 6946 and interpreted this as evidence for a 
correlation between molecular hydrogen and cold dust.
%Recent SCUBA stuff
The relation between gas and dust has been explored using the SCUBA array 
and synthesis mapping of the atomic hydrogen (HI).  Recent SCUBA results \citep{Stevens05,Thomas04} are presented in relation to HI column density. %, both the ratio between gas and dust mass \cite{Stevens05}.
\cite{Stevens05} gives the ratio between gas and dust masses and \cite{Thomas04} 
compare the radial extent of dust and HI and find similar scalelenghts for both. 

%%%%%Reference for the SED, dust and young stars relation...
Spectral energy distribution modeling (SED) of edge-on galaxy disks \citep{Popescu00,Misiriotis01,Popescu02} indicate that the dust's emission is 
powered by HII regions in the case of warm dust ($\lambda < 100 ~ \mu m.$) 
and diffuse stellar radiation in the case of cold dust ($\lambda > 100 ~ \mu m.$).
%the illumination of clouds by a young population of stars. 
The infrared flux is dominated by dust illuminated by nearby stars and an cold dust is 
only visible in the sub-mm part of the spectrum.
Therefore, a cold dust component is relatively more prominent 
at higher radii where the dust clouds are not illuminated as much by the stellar population.

%EXTINCTION MEASUREMENTS
Parallel to this observational effort to characterize the emission, there 
is an effort to characterize the extinction in disks using known background sources. 
Two types of known background sources are in use: occulted galaxies and the number 
of distant galaxies. The occulting galaxy technique has been exhausted on the rare 
nearby pairs \citep{kw92,Andredakis92,kw99a,kw00a,kw00b,kw01a,kw01b}. \cite{kw99a} 
compared the extinction and dust emission in their pairs and found reasonable agreement 
between dust masses and no need for an extremely cold ($\rm T < 10 K.$) component. 

The present paper is one of a series using the number of distant galaxies seen through 
the foreground disk in Hubble Space Telescope (HST) images as an extinction probe. 
The identified number of distant galaxies suffers from crowding and confusion effects. 
To calibrate this observed number, \cite{Gonzalez98} developed the ``Synthetic Field 
Method'' (SFM). A series of synthetic fields is constructed, with the original science field 
to which a dimmed deep field is added. From the relation between detected added 
background galaxies and the dimming of the deep field, the average dimming of the 
science field can be inferred. \cite{Holwerda05a} automated this method and 
\cite{Holwerda05b} reported on the radial opacity profiles of a sample of nearby galaxies.

% What we are going to do here...
In this paper, the radial opacity profiles from \cite{Holwerda05b} are compared to 
HI surface density profiles from the literature for the same galaxies. The inferred 
ratios and radial profiles are compared to those obtained from sub-mm observations 
and others in the literature. 

This paper is organized as follows: section \ref{secHISFM} gives a brief description 
of the ``Synthetic Field Method''. The HI surface density radial profiles, radial opacity 
profiles and their ratios are presented in section \ref{secAHI}.
Radial opacity-to-HI profiles averaged for the whole sample are compared to values 
from the literature in section \ref{secRAHI}. Section \ref{secSCUBA} compares the 
sub-mm profiles from the literature to our opacity profiles. The conclusions are 
summarized in section \ref{secHIconcl} with a view to future work in section \ref{secHIfut}.

\section{\label{secHISFM}The ``Synthetic Field Method''}

The number of distant galaxies seen through a spiral disk does not only depend on the level 
of extinction in the disk but also on the crowding and confusion by the objects in the 
foreground disk. To calibrate the effects of crowding and confusion, \cite{Gonzalez98} 
and \cite{Holwerda05a} developed the ``Synthetic Field Method'' (SFM). This method 
consists of several steps. First the number of distant galaxies seen in the science field 
is identified. The selection is based on object characteristics and color and visually 
checked. Secondly, a series of synthetic fields is constructed. These are the original 
science field with a  Hubble Deep Field added, dimmed to mimic dust extinction. 
Thirdly, the added distant galaxies are identified in these synthetic fields. 
A relationship between the dimming of the synthetic field (A) and the number of added 
galaxies retrieved (N) can be found and we fit the following equation to this:

\begin{equation}
\label{eq:HIAN}
A = -2.5 \ C\ log \left({N \over N_0 }\right),
\end{equation}

\noindent where $C$ characterizes the crowding and confusion for this particular 
science field and $N_0$ is the number of galaxies expected in the case of no dimming 
by dust. Substituting $N$ by the actual number of galaxies found in the science field, 
equation \ref{eq:HIAN} yields the average opacity for the field. The counts of distant 
galaxies are done in I band images and hence we report the opacities as $A_I$.

Cosmic variance in the number of distant galaxies in a given field adds an 
extra uncertainty to the number of distant galaxies found in the science field. As a 
result, opacity measurements in a single WFPC2 field or section thereof have 
high uncertainties associated with them. To combat this, we have automated this 
method \citep{Holwerda05a} and applied it to a sizeable sample of archival WFPC2 
fields \citep{Holwerda05b}. In this series of papers, we have explored the relations 
of disk opacity with radius \citep{Holwerda05b}, the surface brightness \citep{Holwerda05e} 
and HI in this paper. \cite{Holwerda05d} explore the limitations of this method as 
predicted by \cite{Gonzalez03}, concluding that the optimal foreground disk distance 
is somewhere between 5 and 30 Mpc.

\section{\label{secAHI}Individual galaxies: radial HI and opacity profiles}

The number of distant background galaxies as a function of HI column density is 
best directly measured using an overlay of the HI column density map on an HST 
field. See for example the analysis by \cite{Cuillandre01} of a ground-based field 
in M31.  However, since the HI column density maps are not easily available and 
not uniform, the relation between radial profiles of HI and opacity is used.

\cite{Holwerda05b} present radial profiles for individual galaxies and composites 
of fields (their Table 3). HI surface density profiles were taken from the literature 
for the subset of our sample for which these were available (Table \ref{HItable}). 
HI surface density profiles of the galaxies were extracted from the literature using 
the DEXTER program \citep{dexter} and rescaled to express radius in $R_{25}$ 
\citep{RC3} and surface density in ${\rm M_{\odot} ~ pc^{-2}}$. 

In Figure \ref{RAHI1} and \ref{RAHI2} the individual radial surface density profiles 
of HI and the opacity profiles are shown. The opacity was determined for the 
sections of the WFPC2 fields corresponding to radial intervals of 0.25 $R_{25}$.
Both the opacity profiles and the HI surface density profiles of spiral disks display 
a variety of shapes. However, in general, the HI profile peaks somewhere in 
the disk and flattens out or dips near the galaxy's center. The opacity profiles show 
a gradual rise towards the disk's center. It should be noted that the HI profiles are the 
azimuthally average for the entire disk, while the opacity profile is derived from the 
smaller section of the disk corresponding to the HST image i.e. these profiles are for 
the same spiral galaxy but not determined from the same section of the disk.

To see if there is correlation between HI surface density and disk opacity, the values 
of the profiles in Figures \ref{RAHI1} and \ref{RAHI2} are plotted in Figure \ref{AHI}.
Averaged over radial intervals of 0.25 $R_{25}$, there seems to be no correlation 
between HI column density and dust opacity in a spiral disk. The lack of a relation 
may be explained by the fact that the opacity profiles generally rise in the center of 
galaxies  \citep{Holwerda05b}, while radial HI surface density profiles often show a 
drop in the galaxy's center (See also Figures \ref{RAHI1} and \ref{RAHI2}). Alternatively, 
part of the hydrogen gas may be associated in molecular clouds, undetected in HI 
observations.

\section{\label{secRAHI}Dust-to-HI ratio}

To reduce the uncertainties in the opacity measurement, the galaxy counts from 
several fields must be combined. This can be done e.g. per Hubble type or for the 
entire sample (See \cite{Holwerda05b} for this type of analysis). 

In Figure \ref{AV}, the average HI-to-opacity plot based on our entire sample 
(Table \ref{HItable}) is compared to HI-to-opacity measurements from the 
literature. To obtain this HI-to-dust profile, the opacities were re-derived and the 
HI profiles averaged. First, all the counts of distant galaxies, from all the synthetic 
and science fields, were combined. Then the opacity was derived again from the 
combined counts in radial intervals, using  equation \ref{eq:HIAN}. The HI surface 
density profiles from figures \ref{RAHI1} and \ref{RAHI2} were combined by adding 
mass and surface contributions of each profile in a radial section and taking the ratio 
of the sums.

% How does our profile look?
An average opacity taken over many galaxies does not suffer from poor statistics 
as the individual profiles do, but small scale variations are smoothed out. The radial 
profile of dust-to-HI remains relatively constant for most of the optical disk and 
turns upward beyond that. At this point, the HI profiles diminish but the opacity 
profiles remain more or less constant. 
The radial profile becomes more uncertain beyond $R_{25}$ as the WFPC2 
fields used for the SFM analysis were centered on the optical disk (See for more 
on the selection of our original sample \cite{Holwerda05b}.). For illustration, 
only two galaxies have enough counts for a reasonable SFM measurement 
beyond the $R_{25}$. The trend of the dust-to-HI ratio beyond the $R_{25}$ 
is therefore correspondingly uncertain. 

% How was the comparison data obtained?
The comparison data from the literature are based on a series of techniques to 
determine the dust content. 
\cite{Issa90} compile dust-to-HI measurements from the literature for a few nearby 
galaxies and normalize these to the Galactic value of \cite{Bohlin78} and a distance 
of 0.7 $R_{25}$ from their galaxies's center. These values from \cite{Issa90} are 
derived from the line ratio between H$\alpha$ and H$\beta$. 
\cite{Mayya97b} infer average dust-to-HI profiles from IRAS emission at 60 micron 
which traces only the warm dust. They give a range of conversion values for the optical 
depth at 60 micron and in V. 
\cite{Boissier04} present radial dust-to-HI profiles for 6 galaxies deduced from the 
FIR/UV ratio. 
\cite{Cuillandre01} derive a relation between HI column density and stellar reddening 
for M31. The average HI-to-dust ratio from the sub-mm observations of disks by 
\cite{Stevens05} is also indicated. Their total sub-mm flux density was 
converted using the ratio found by \cite{Alton98a} and an estimate of the disk area using 
the $D_{25}$.

% Why the difference?
The comparison to other results in Figure \ref{AV} raises the question why the opacity 
measurement from the number of distant galaxies results in a dust-to-HI ratio that is 
an order of magnitude higher than the nearest recent estimate \citep{Boissier04}. 
%
% Our possible problems
There are several effects which could have affected the average dust-to-HI profile 
from counts of galaxies.
(1) The opacity measure from the counts of galaxies is averaged over a series of Hubble 
types and HI profiles. 
% HI profiles underestimate HI column density?
(2) The older HI surface density profiles may underestimate the HI column density, 
especially the larger scale structure due to lack of short-spacing information.
% Opacity values overestimates
%% Arm
(3) The opacities are predominantly determined for fields with a spiral arm while the HI 
profiles are averages over the whole of disk. This may have lowered the HI column 
density profile with respect to the measured opacity.
(4) The average opacity measurement from counts of galaxies indicates the covering 
factor of dark clouds. One could argue that this covering factor is an overestimate as 
the distant galaxy does not need to be completely covered in order to be dropped 
from the counts. All these effects could have lowered the dust-to-HI profile presented here. 

% Other people's problems
The other profiles also suffer from various different systematics.
The dust-to-HI ratio estimates from the literature rely on the light of the disk itself to 
estimate the dust content. This inherently biases the measurement to lower extinction 
values, as not the entire height of the dust disk is responsible for the observed extinction 
and the measured light is biased towards low-extinction lines of sight. The different 
manner in which the dust content was estimated may also play a role.
The values from \cite{Issa90} are for very specific parts of the disk, e.g. the HII regions, 
and these were scaled to a single radius. Still, the ratio is similar to those found by 
other authors for the whole of the disk. The infrared measures of \cite{Mayya97b} 
($\lambda = 60 ~ \mu m.$) show substantially more HI for a given amount of dust. 
Most likely this is the effect of their observational technique which is more sensitive 
to the warmer component of the dust in the spiral disk, which dominates the infrared 
emission. The fact that infrared emission only detects 10 to 20 \% of the dust in a 
disk was also found by \cite{Devereux90} from 60 and 100 $\mu m.$ emission 
observed with IRAS.
By taking the ratio of UV and far-IR flux, \cite{Boissier04} can characterize better 
the total dust content of a disk. The average value from \cite{Stevens05} is again closer 
to this paper's profile. 

The progression from the values of \cite{Mayya97b} to those of \cite{Boissier04} and 
\cite{Stevens05} indicates that the more sensitive the dust indicators become for the 
colder dust component in a disk, the more dust is detected in relation to the HI. It has also been 
found by several authors \citep{Alton98a,Popescu02} that the bulk of the dust reveals 
itself when more sensitive dust indicators are used. Taken in this context, the SFM 
estimate (the thick line in Figure \ref{AV}) of dust content of disks, which is independent 
of dust temperature, may reflect the actual total dust-to-HI ratio. 
 
\section{\label{secSCUBA}Comparison to SCUBA profiles.}

The SCUBA detector on the JCMT detects sub-mm emission from cold dust in spiral 
disks. In order to verify whether or not this cold dust is responsible for the opacity 
measured with the SFM, a direct comparison should be made. Four galaxies in 
our sample were mapped in the 850 $\mu m$ band by different authors 
\citep{Alton01b,Alton02,Meijerink05,Stevens05}. However, the individual opacity 
measurements and the conversion from 850 micron flux to an optical depth do not 
allow for a good comparison. The exact emissivity of dust grains at these wavelengths 
may still be underestimated \citep{Alton00b, Alton04,Dasyra05}. To illustrate a 
comparison between a SCUBA map and our counts, we present the best current 
example: the exponential disk found by \cite{Meijerink05} for M51 and our distant 
galaxy counts from two WFPC2 fields.

\subsection{\label{ssecM51scuba}M51}

\cite{Meijerink05} present a detailed map of the 850 micron emission, as well as an exponential
optical depth profile for the dust disk. Figure \ref{M51scuba} shows the 
850 micron emission profile from \cite{Meijerink05} translated to an opacity value 
in V and I and the opacity measurements of the two WFPC2 fields and their average.
Our WFPC2 fields predominantly miss the sections of the disk for which \cite{Meijerink05} 
report additional flux from the spiral arms. However, a proper comparison can be done 
when the sub-mm observations and the counts of distant galaxies are from a similar sized field.
For now we can only conclude that the points and the profile seem to agree.

\subsection{Comparing to sub-mm}

A good comparison between opacity and sub-mm flux can be obtained with more and better 
sub-mm maps and sufficient counts of distant galaxies. The new SCUBA-2 instrument on the 
JCMT promises to facilitate this mapping of nearby galaxies. However the expected emissivity 
of dust grains must then also be known for an accurate comparison.
An advantage of the counts of distant galaxies is that they can be extended to larger radii 
and lower opacity values and can independently verify the profile found from sub-mm observations. 

Both observational techniques can be used to explore the radial extent of dust in spiral disks. 
Our previous composite profile \citep{Holwerda05b} already pointed to and extended dust 
disk as well as several sub-mm observations \citep{Nelson98, Alton98,Trewhella00,Popescu03}. Currently the extent is taken to be somewhere between the optical and the HI scale. A more 
accurate determination of that can be achieved with either more sensitive sub-mm observations 
or future galaxy counts or both combined.

\section{\label{secHIconcl}Conclusions}

In summary, we draw the following conclusions from the comparison between opacity from the 
number of distant galaxies with HI profiles:
\begin{itemize}
\item[1.] The HI and opacity profiles of individual fields do not seem to correlate well. This may 
be due to the rise of opacity in the center of a disk while there is often a dip in the atomic hydrogen surface density profile there. However, high uncertainties plague the single field opacity 
measurements (Figures \ref{RAHI1}, \ref{RAHI2} and \ref{AHI}).
\item[2.] The dust-to-HI ratio depends on the tracer used for the dust content. When a tracer 
sensitive to colder dust is used, a higher dust-to-HI value is found (Figure \ref{AV}). The SFM 
opacity, which is independent of the dust temperature, could very well point to the true dust to 
HI ratio in spiral disks.
\item[3.] A direct comparison between sub-mm emission and SFM opacity is problematic as 
few galaxies in our sample have been observed at these wavelengths. It is possible that both 
methods trace the same component (Figure \ref{M51scuba}) but more data and a better 
understanding of dust emissivity is needed for confirmation. 
\end{itemize}

\section{\label{secHIfut}Future Work}

% Future sub-mm work
The relation between sub-mm emission and opacity is an important area for future work. A more accurate comparison between the SFM opacity profile and the SCUBA emission profile of M51 should become possible in the near future with the HST/ACS observations by the {\it Hubble Heritage} Team. A similar comparison could be made for M101, provided such a large solid angle could be successfully scanned by the new SCUBA-2 or its successors.

%Future HI work
By comparing the SFM opacity to HI column density averaged over a radial interval, small variations are smoothed out. A more direct approach is to compare SFM opacity directly in contours of HI column density. This requires uniform HST imaging of a nearby galaxy and a deep and uniform HI map as well. Such data exists for M101 and M51 and the relation between HI and dust could then be characterized more accurately and out to a larger radius.

\acknowledgements
The authors would like to thank the referee, Simone Bianchi for his comments, 
Jason Stevens for making the SCUBA map of NGC 4414 available 
and George Bendo for useful discussions on SCUBA observations.
This research has made
use of the NASA/IPAC Extragalactic Database, which is operated
by the Jet Propulsion Laboratory, California Institute of
Technology, under contract with the National Aeronautics and
Space Administration (NASA). This work is primarily based
on observations with the NASA/ESA Hubble Space Telescope,
obtained at the STScI, which is operated by the Association of
Universities for Research in Astronomy (AURA), Inc., under
NASA contract NAS5-26555. Support for this work was provided
by NASA through grant number HST-AR-08360 from
STScI to Prof. Dr. R.J. Allen. STScI is operated by AURA, Inc., under NASA contract
NAS5-26555. We are also grateful for the financial support
of the STScI DirectorÕs Discretionary Fund (grants 82206 and
82304 to R. J. Allen) and of the Kapteyn Institute of Groningen University.

%\bibliographystyle{aa} 
%\bibliography{./bib/HIprofiles,./bib/paperII,./bib/paperIII,./bib/paperIV,./bib/Alton,./bib/Boissier_ref,./bib/Fischera,./bib/GasToDust,./bib/Maiolino,./bib/Holwerda,./bib/SCUBA,./bib/Greeks}

\newpage

%%%%%%Table 1%%%%%%
\begin{deluxetable}{l l l l l l l l l l l}
\tablewidth{0pt}
\tablehead{
\colhead{Galaxy} 	& 
\colhead{Hubble Type} 	& 
\colhead{Instrument} 	& 
\colhead{$R_{25}$}		&
\colhead{Reference} 	\\
	& 	& 	& arcmin.	& \\}

\startdata
			
NGC 925			& SAbd		& WSRT 	& 10.47	& \cite{Wevers86}\\
				& 			& VLA 	&		& \cite{Pisano98}\\ 
NGC 1365		& SBb     		& VLA  	& 11.22	& \cite{Jorsater95}\\ 
NGC 1425		& SBb		& - 		& 5.75	& - \\
NGC 1637		& SAB(rs)c 	& - 		& 3.98	& - \\
NGC 2541		& SAcd    		& WSRT	& 6.31	& \cite{Broeils94} \\
NGC 2841		& SAb     		& WSRT 	& 8.13	& \cite{Bosma81a} \\
NGC 3198		& SBc     		& WSRT 	& 8.51	& \cite{Wevers86} \\ 
NGC 3319		& SB(rs)cd 	& WSRT 	& 6.17	& \cite{Broeils94} \\
NGC 3351		& SBb		& - 		& 7.41	& - \\			
NGC 3621		& SAc 	  	& - 		& 12.3	& - \\		
NGC 3627		& SAB(s)b		& - 		& 9.12	& - \\		
NGC 4321		& SABbc   	& WSRT	& 7.41	& \cite{Warmels88c}\\
				&			& VLA 	&		& \cite{Cayatte94} \\
NGC 4414		& SAc     		& VLA 	& 3.63	& \cite{Thornley97}\\
NGC 4496A		& SBm	  	& - 		& 3.98	& - \\
NGC 4527		& SAB(s)bc	& - 		& 6.17	& - \\			
NGC 4535		& SABc     		& WSRT 	& 7.08	& \cite{Warmels88c}\\
				& 			& VLA 	&		& \cite{Cayatte94}\\	
NGC 4536		& SAB(rs)bc 	& - 		& 7.59	& - \\					
NGC 4548	 M91	& SBb		& WSRT	& 5.37	& \cite{Warmels88c}\\
				&			& VLA 	&		& \cite{Cayatte94}\\
NGC 4559		& SAB(rs)cd	& WSRT 	& 10.72	& \cite{Broeils94}\\ 
NGC 4571		& SA(r)d  		& WSRT 	& 3.63	& \cite{Warmels88c}\\		
NGC 4603		& SA(rs)bc 	& - 		& 3.39	& - \\
NGC 4639		& SABbc	  	& WSRT 	& 2.75	& \cite{Warmels88c}\\		
NGC 4725		& SABab	  	& WSRT 	& 10.72	& \cite{Wevers86}\\ 
NGC 6946		& SAB(rs)cd	& VLA-D 	& 11.48	& \cite{Tacconi86}\\ 
NGC 7331		& SAb	  	& WSRT 	& 10.47	& \cite{Bosma81a}\\		
M51				& SA(s)bc 	& WSRT 	& 5.61	& \cite{Bosma81a}\\		
M81				& SA(s)ab		& WSRT 	& 13.46	& \cite{Cayatte94}\\
\enddata
\tablecaption{HI surface density profiles from the literature. Some of these profiles may suffer from short-spacing problems which would underestimate the flux from HI on larger scales.}
\label{HItable}
\end{deluxetable}
\clearpage
\newpage

%------------------------------- FIGURE 1a -------------------------------
\begin{subfigures}

\begin{figure}
\includegraphics[width=0.5\textwidth]{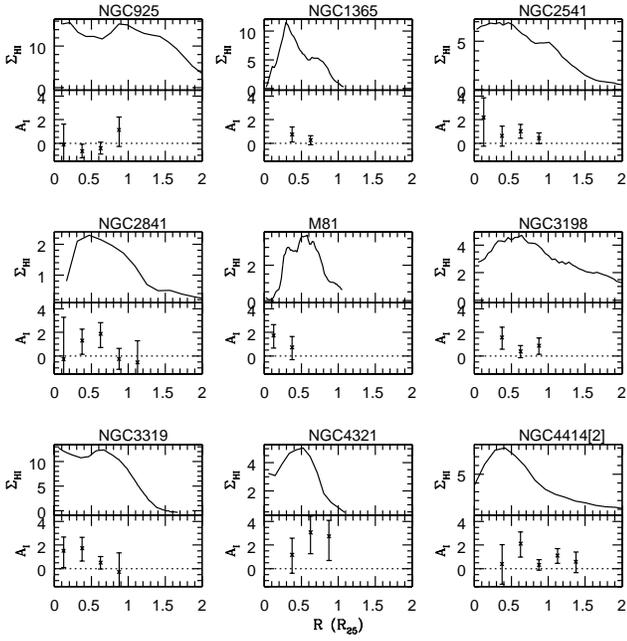}
\caption{\label{RAHI1}The HI and opacity profiles of individual galaxies in our sample. $A_I$ is expressed in magnitudes, the atomic hydrogen surface density ($\rm \Sigma_{HI}$) in solar masses per square parsec ($\rm M_{\odot} pc^{-2}$). The HI surface density profiles often do not have uncertainties reported with them. See Table \ref{HItable} for the original reference for the HI surface density profiles. The most recent profiles are used in all cases. NGC 4414 has two WFPC2 fields associated with it.}
\end{figure}

%------------------------------- FIGURE 1b -------------------------------
\begin{figure}
\includegraphics[width=0.5\textwidth]{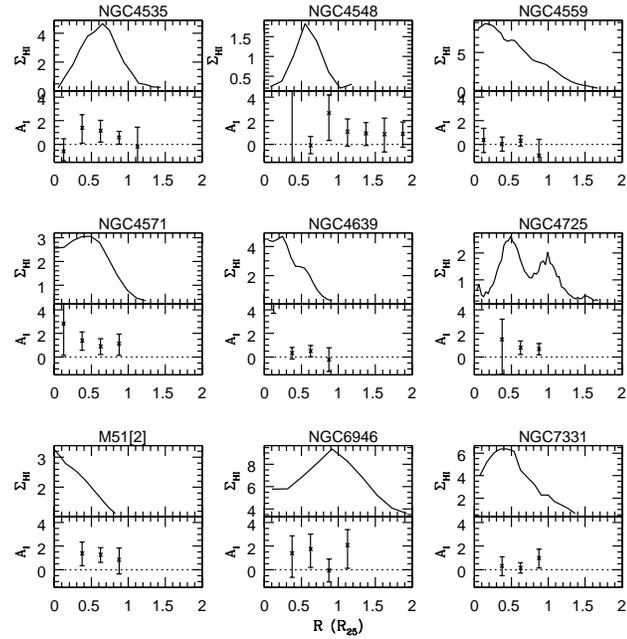}
\caption{\label{RAHI2}The HI and opacity profiles of individual galaxies in our sample. }
\end{figure}
\end{subfigures}

%------------------------------- FIGURE 2 -------------------------------
\begin{figure}
\includegraphics[width=0.5\textwidth]{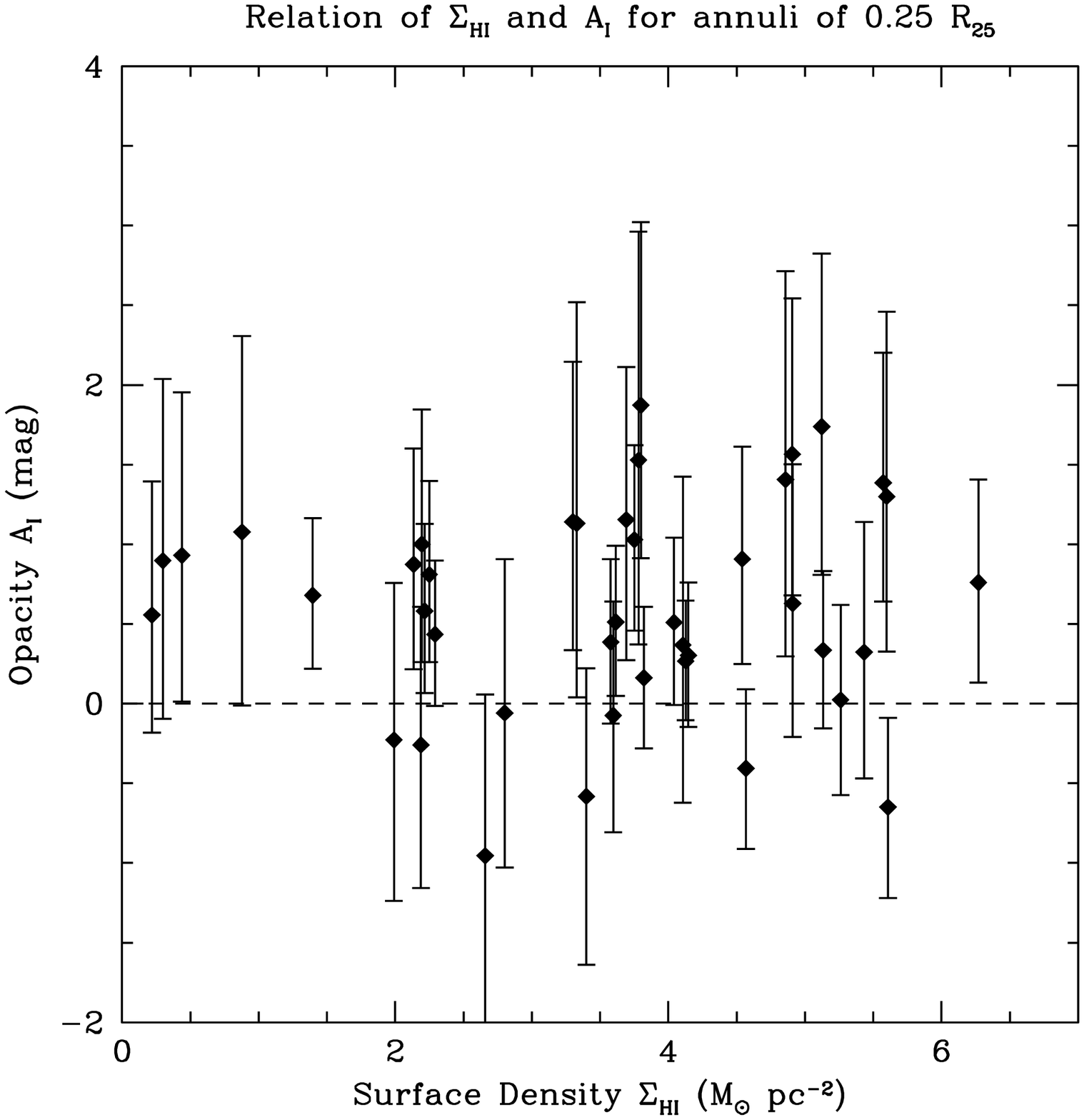}
\caption{\label{AHI}The average opacity ($A_I$) and surface density ($\Sigma_{HI}$) for radial annuli of 0.25 $R_{25}$ in individual galaxies in our sample (See Figures \ref{RAHI1} and \ref{RAHI2}).}
\end{figure}

%------------------------------- FIGURE 3 -------------------------------
\begin{figure}
\includegraphics[width=0.5\textwidth]{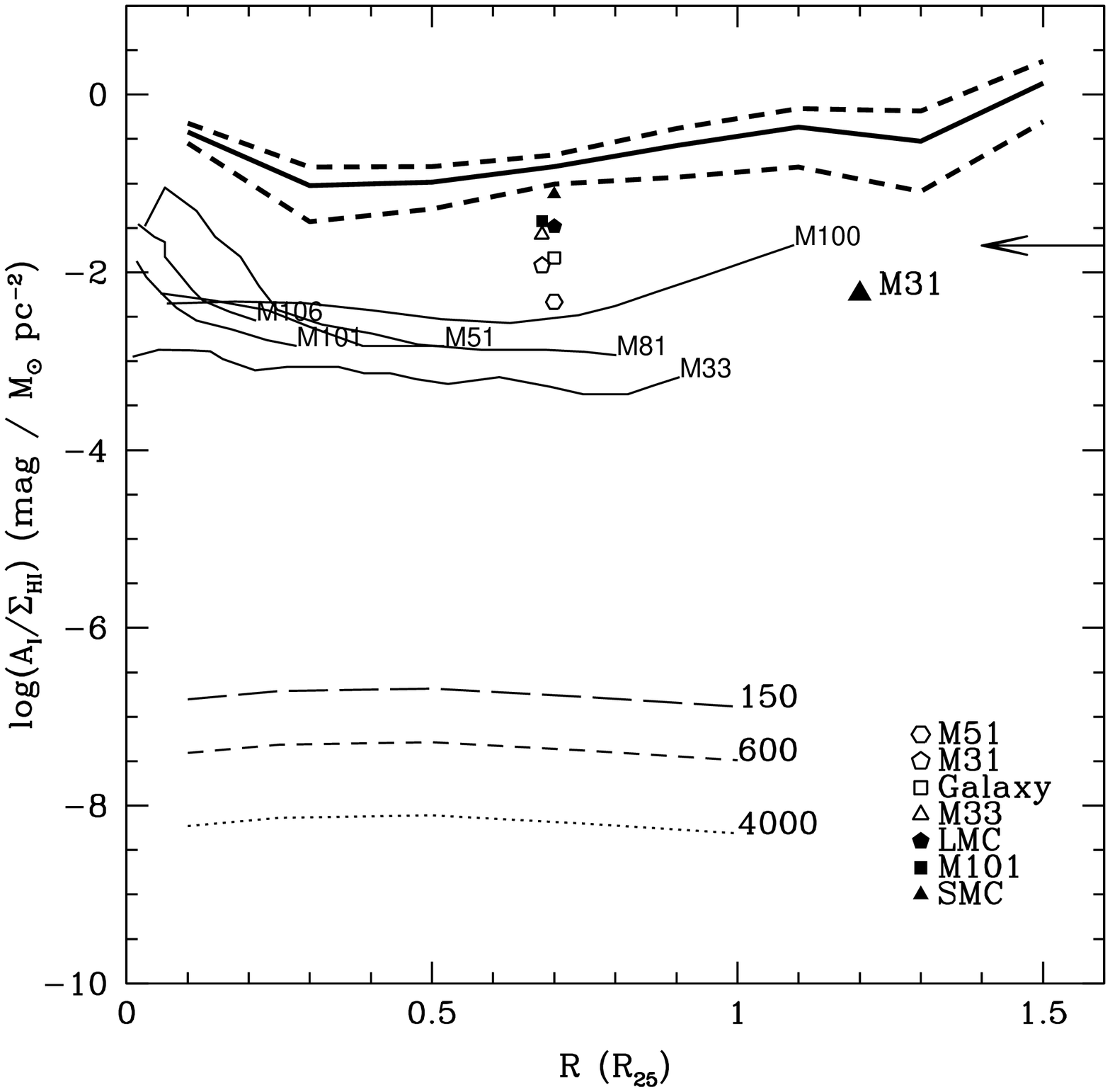}
\caption{\label{AV}The radial dust-to-HI profiles from \cite{Issa90}, \cite{Mayya97b}, 
\cite{Boissier04} and this paper (thick line, uncertainties are dashed).
The symbols are the values from \cite{Issa90}(H$\alpha$/H$\beta$ ratio). 
Note how these are higher for the same galaxy as the values from 
\cite{Boissier04}.
The three bottom lines are the \cite{Mayya97b} (IRAS 60 $\mu m$ emission) 
average ratio, for the different conversion factors between the 60 $\mu$m 
optical depth and the V band optical depth ($\tau_V ~ / ~ \tau_60$). 
The curves in the middle are the dust-to-HI profile from \cite{Boissier04}(UV/FIR ratio), 
for their 6 galaxies, converted to I band extinctions using \cite{Boselli03}. 
The arrow right is the average ratio from \cite{Stevens05} (850 $\mu m$ emission), 
the combined warm and cold dust over HI. }
\end{figure}

%------------------------------- FIGURE 4 -------------------------------
\begin{figure}
\includegraphics[width=0.5\textwidth]{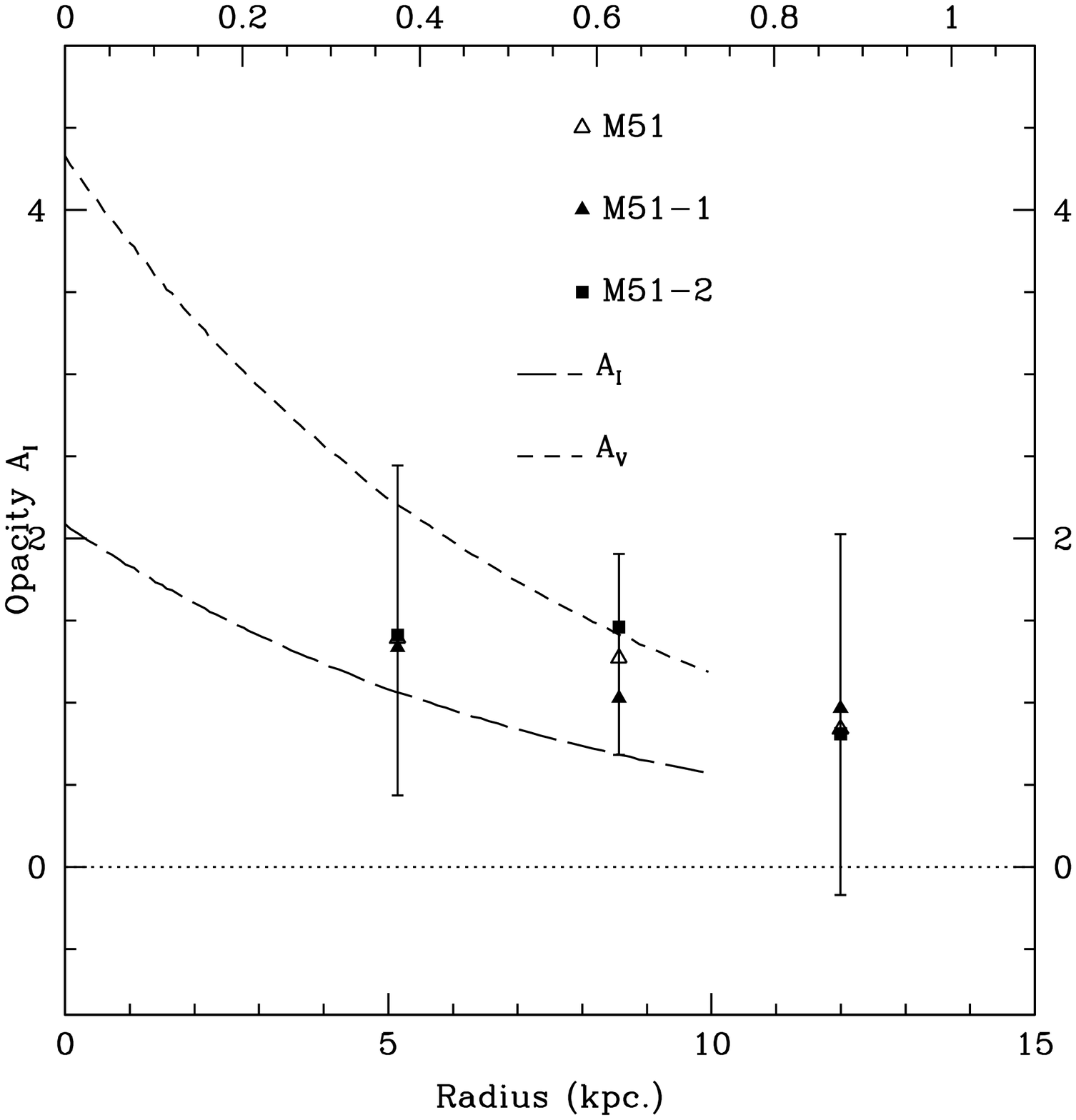}
\caption{\label{M51scuba}The profile of M51 from \cite{Meijerink05} (dashed lines) 
and the opacities from \cite{Holwerda05b} (points). Solid points are the individual 
WFPC2 fields, the open triangles are the derived from combined WFPC2 fields. 
The errorbars shown are for the combined opacity and the radial bin size is 0.25 
$R_{25}$. The top axis shows the radius in $R_{25}$). 
Conversion from the SCUBA profile to an opacity in V used two times the Galactic emissivity at 850 micron and the Galactic reddening law was used to convert the opacity profile in V to one in I.}
\end{figure}

\end{document}